\documentclass[12pt]{article}
\usepackage{amsmath}
\usepackage{amsfonts}
\usepackage{amssymb}
\usepackage{amsthm}

\usepackage[unicode=true,pdfusetitle,
 bookmarks=true,bookmarksnumbered=false,bookmarksopen=false,
 breaklinks=true,pdfborder={0 0 0},backref=false,colorlinks=true]
 {hyperref}
\usepackage{mathrsfs}
\usepackage{subcaption}
\usepackage{cancel}

\usepackage{color}
\newcommand{\bc}{\color{black}}

\newcommand{\bcol}{\color{black}}

\newlength{\dinwidth}
\newlength{\dinmargin}

\newif\ifshowdetails
\showdetailstrue
\hypersetup{
 linkcolor=blue,citecolor=blue, urlcolor=blue}

\newcounter{todo}

\newcommand{\todobox}[1]{
  \textcolor{blue}{
    \fbox{\parbox{0.9\textwidth}{#1}}%
  }
}

\newcommand{\todonotetag}{TODO\thetodo}
\newcommand{\todonote}[1]{%
\stepcounter{todo}%
{\let\thefootnote\todonotetag
\footnote{\todobox{#1}}%
}
}

\definecolor{detailsgray}{gray}{0.3}

\ifshowdetails
\newcommand{\detail}[1]{%
{\color{detailsgray}$\blacktriangleright${#1}$\blacktriangleleft$}%
}

\newcommand{\detailspar}[1]{
\par \noindent {\color{detailsgray} $\blacktriangleright$ \textit{#1} $\blacktriangleleft$ } \par
}

\newenvironment{details}{\par \noindent \begingroup\itshape\color{detailsgray} $\blacktriangleright$ \hrulefill%
\par \noindent \ignorespaces}{\par\noindent \hrulefill $\blacktriangleleft$  \endgroup \par \noindent}

\else
\newcommand{\detail}[1]{} 
\newcommand{\detailspar}[1]{} 
\usepackage{environ}
\NewEnviron{hide}{}

\fi

\newcommand{\free}{{\bc\mrm{fr}}}
\newcommand{\vn}{\mathbf{n}}
\newcommand{\vy}{\mathbf{y}}
\newcommand{\vB}{\mathbf{B}}
\newcommand{\vr}{\mathbf{r}}

\newcommand{\vE}{\mathbf{E}}
\newcommand{\ve}{\mathbf{e}}
\newcommand{\vh}{\mathbf{h}}

\newcommand{\vfe}{\vf_{\mrm{el}}}
\newcommand{\vfb}{\vf_{\mrm{b}}}

\usepackage{tikz}
\usetikzlibrary{trees}
\usetikzlibrary{decorations.pathmorphing}
\usetikzlibrary{decorations.markings}

\newcommand{\hvk}{\hat{\vk}}

\newcommand{\vv}{\pmb{v}}

\newcommand{\mrm}{\mathrm}

\newcommand{\vA}{\mathbf{A}}

\newcommand{\vf}{\mathbf{f}}

\newcommand{\veps}{\epsilon}

\renewcommand{\mathbf}{\boldsymbol}

\newcommand{\mcF}{\mathcal F}

\newcommand{\mcL}{\mathcal L}

\newcommand{\E}{E}

\newcommand{\wt}{\widetilde}

\newcommand{\vx}{\boldsymbol{x}}

\newcommand{\ti}{\tilde}

\newcommand{\Om}{\Omega}

\newcommand{\vk}{\boldsymbol{k}}

\newcommand{\pa}{\partial}

\newcommand{\ov}{\overline}

\newcommand{\hk}{\hat{k}}
\newcommand{\eps}{\varepsilon}
\newcommand{\de}{\delta}
\newcommand{\De}{\Delta}

\newcommand{\nin}{\noindent}
\newcommand{\si}{\sigma}
\newcommand{\ph}{\phantom}
\newcommand{\h}{\fr{1}{2}}

\newcommand{\om}{\omega}
\newcommand{\mfa}{\mathfrak{A}}
\newcommand{\mco}{\mathcal{O}}

\newcommand{\supp}{\mathrm{supp}}
\newcommand{\fr}[2]{\frac{#1}{#2}}

\newcommand{\real}{\mathbb{R}}
\newcommand{\complex}{\mathbb{C}}
\newcommand{\la}{\lambda}
\newcommand{\non}{\nonumber}

\newcommand{\lan}{\langle}
\newcommand{\ran}{\rangle}

\def\qed{$\Box$\medskip}
\newtheorem{theoreme}{Theorem } [section]
\newtheorem{proposition}[theoreme]{Proposition}
\newtheorem{lemma}[theoreme]{Lemma}
\newtheorem{definition}[theoreme]{Definition}
\newtheorem{corollary}[theoreme]{Corollary}
\newtheorem{remark}[theoreme]{Remark}
\newtheorem{example}[theoreme]{Example}
\newtheorem{criterion}[theoreme]{Criterion}
\newtheorem{conjecture}{Conjecture}
\newtheorem{assumption}{Assumption}

\newcommand{\bea}{\begin{assumption}}
	\newcommand{\eea}{\end{assumption}}
\newcommand{\beco}{\begin{conjecture} }
	\newcommand{\eeco}{\end{conjecture} }
\newcommand{\beq}{\begin{equation}}
	\newcommand{\eeq}{\end{equation}}
\newcommand{\beqa}{\begin{eqnarray}}
	\newcommand{\eeqa}{\end{eqnarray}}
\newcommand{\ben}{\begin{arabicenumerate}}
	\newcommand{\een}{\end{arabicenumerate}}
\newcommand{\bex}{\begin{example}}
	\newcommand{\eex}{\end{example}}
\newcommand{\ber}{\begin{remark}}
	\newcommand{\eer}{\end{remark}}
\newcommand{\bec}{\begin{corollary}}
	\newcommand{\eec}{\end{corollary}}
\newcommand{\bep}{\begin{proposition}}
	\newcommand{\eep}{\end{proposition}}
\newcommand{\becr}{\begin{criterion}}
	\newcommand{\eecr}{\end{criterion}}

\def\bel{\begin{lemma}}
	\def\eel{\end{lemma}}
\def\bet{\begin{theoreme}}
	\def\eet{\end{theoreme}}
\def\bed{\begin{definition}}
	\def\eed{\end{definition}}

\title{Asymptotic charges, large gauge transformations and inequivalence of different gauges in external current QED}

\author{
{\bf Wojciech Dybalski\footnote{This work was  supported by the DFG within the Emmy Noether grant DY107/2-1.} }\\
Zentrum Mathematik, Technische Universit\"at M\"unchen,\\
E-mail: {\tt dybalski@ma.tum.de}
\and
{\bf Benedikt Wegener}\\
Marie Sklodowska-Curie fellow of the Istituto Nazionale di Alta Matematica\footnote{This project has received funding from the European Union's 2020 research and innovation programme under the Marie Sklodowska-Curie grant agreement No 713485.}\\
Dipartimento di Matematica, Universit\`a di Roma ``Tor
Vergata''\\ 
E-mail: {\tt wegener@mat.uniroma2.it}}

 \date{}

\begin{document}

\maketitle

\begin{abstract}
In this paper we consider external current QED in the Coulomb gauge and in axial gauges
for various spatial directions of the axis.  For a non-zero electric charge of the current, we demonstrate that any two different  gauges from this class
correspond to quantum theories which are  not unitarily equivalent. We show that the spacelike asymptotic flux of the electromagnetic field
is the underlying superselected quantity. We also express the large gauge transformation linking two distinct
axial gauges by the Wilson loop over a contour limited by the two axes. 
Thus the underlying physical mechanism appears to be related to the Aharonov-Bohm effect.
\end{abstract} 
%\vspace{0.2cm}


%\vspace{0.2cm}

%%%%%%%%%%%%%%%%%%%%%%
\section{Introduction}
%%%%%%%%%%%%%%%%%%%%%%
\setcounter{equation}{0}

In Classical Electrodynamics  a change of gauge of the electromagnetic
potential $A_{\mu}\mapsto A_{\mu}+\pa_{\mu}\chi$ has clearly no observable
effect as it does not change the electromagnetic
fields $F_{\mu\nu}$.  This argument does not extend to Quantum Electrodynamics,  since
the quantization procedures always require some gauge fixing conditions. In spite of 
the seemingly unlimited gauge freedom, the list of conditions used in practice is relatively
short: the Lorentz gauge $\pa_{\mu}A^{\mu}=0$, the Coulomb gauge $\nabla\cdot  \pmb{A}=0$
and axial gauges $e_{\mu} A^{\mu}=0$ for various directions of the axis $e$ are the most common 
choices. {\bc Some} authors have argued that the resulting theories are equivalent  on the basis
of partially heuristic computations \cite{HL94,Na94}. In the present paper we consider external current QED in 
the Coulomb  and axial   gauges and show that actually the opposite is the case:  the corresponding quantum theories are not unitarily equivalent if the current has a non-zero electric charge. {\bc Our analysis is `ghost-free', the results are mathematically rigorous, and  we  exhibit a physical mechanism
behind the inequivalence of different gauges}.   
%Our findings are
%not only mathematically rigorous, but also demonstrate a physical mechanism behind the 
%inequivalence of different gauges.

Let us outline this mechanism briefly: It is well known that the {\bc electric} charge conservation follows from the
Noether theorem applied to the global U(1) symmetry. It is less well known that the Noether theorem
applied to the local gauge symmetry gives conservation of the spacelike asymptotic flux of the electric field\footnote{The argument can be found e.g. in https://en.wikipedia.org/wiki/Infraparticle. }
\beqa
\phi(\vn)=\lim_{r\to\infty} r^2\vn\cdot \vE(\vn r),  \textrm{ where }  |\vn|=1, \label{flux}
\eeqa
which commutes with all local observables. 
In an irreducible representation of the  external current QED the flux $\phi$ is thus a scalar function on the unit sphere, 
restricted only by the Gauss Law.   Any choice of this function corresponds to a different sector of the theory \cite{Bu82}.
Not surprisingly, this function inherits the symmetry of the
gauge-fixing condition used in the quantization procedure: In the Coulomb gauge it is spherically symmetric,
while in the axial gauges it is only axially symmetric. Hence, the respective quantum theories are not unitarily
equivalent and the same is true for axial gauges with distinct directions of the axis. However,  we show that
they are related by a Bogolubov transformation, which can be expressed  by a Wilson loop over a  contour limited by the two axes.   Or,  equivalently,
by the flux of the magnetic field through the area surrounded by this contour.
This brings to light certain similarity  of our findings with the Aharonov-Bohm effect,  where the phase shift of a particle traveling
along two paths with the same start and end points is given by an analogous expression.
%For other relations between the Gauss Law and the
%Aharonov-Bohm effect we refer to \cite{SDH14} 

As our Bogolubov transformations  change the asymptotic charge  (\ref{flux}),  they
have a non-trivial action at infinity. In this respect they resemble the large gauge transformations
considered recently in the context of the Strominger's `infrared triangle', see e.g. \cite{HMPS14,CL15,CE17,GS16}, 
which, however, are residual transformations of the Lorentz gauge\footnote{The existence of large gauge transformations of this type can be questioned \cite{He17}.}. 
 For transformations between the Lorentz gauge and other  covariant gauges, and their 
automorphic action on the field algebra we refer to \cite[Appendix A]{FS15}. For relations between the Aharonov-Bohm effect and
the Gauss Law at the level of the algebra we refer to \cite{SDH14, Sch19}.
An interesting discussion
of the Gauss Law and the flux of the electromagnetic field in the context of axial gauges was recently given in \cite{BCRV19, RMS19}.
{\bcol In  \cite{RMS19} the flux is computed in axial gauges in low orders of perturbation theory, giving results similar to ours. However,  the external
current discussion from this reference corresponds to the Coulomb gauge in our setting.} Finally, we refer to \cite{DDMS10} 
and references therein for the complementary case of the {\bcol second quantized} Dirac field in an external electromagnetic  field
and its sensitivity to the choice of  gauge. However, the problem of unitary inequivalence of different {\bcol gauge fixing conditions}
is not discussed in the above references and, to our knowledge, it does not have a satisfactory treatment
in the literature.

This paper is organized as follows: In Section~\ref{asymptotic-symmetries} we verify that the Coulomb gauge
and the axial gauge are not unitarily equivalent,  if the external current has a non-zero total charge. We also show the inequivalence of  axial gauges with different
directions of the axis. These results are obtained by computing the flux (\ref{flux}) in different gauges.   
In Section~\ref{large-gauge-transformations} we identify the Bogolubov transformations linking different axial gauges
and express them as Wilson loops. In Section~\ref{conclusions} we summarize our work,  outline briefly the case
of angularly smeared axial gauges   and discuss future directions.

%%%%%%%%%%%%%%%%%%%%%%%%%%%%%
\section{Asymptotic symmetries and inequivalence of different gauges}  \label{asymptotic-symmetries}
%%%%%%%%%%%%%%%%%%%%%%%%%%%%%
\setcounter{equation}{0}

Let us first recall the standard formulas for the free transverse  potential and {\bc free} electromagnetic fields  (see e.g. \cite{Wi95})
\begin{align}
\vA_{\bot}(t,\vx)&:= \fr{1}{(2\pi)^{3/2}}\sum_{\la=\pm }\int \fr{d^3\pmb{k}}{\sqrt{2|\vk|} }\, \pmb{\epsilon}_{\la}(\vk)\big( e^{i|\vk|t-i\vk\cdot \vx}a^*_{\la}(\vk)+   e^{-i|\vk|t+ i\vk\cdot \vx}a_{\la}(\vk)  \big),\\
\vE_{\free}(\vx)&:=-\pa_t  \vA_{\bot}(t,\vx), \quad  \vB_{\free}(\vx):=\nabla \times  \vA_{\bot}(t,\vx),
\end{align}
where $a^{(*)}_{\la}$ are the creation/annihilation operators of  photons on the Fock space $\mcF$ and  $\pmb{\epsilon}_{\la}$ are the polarization vectors.
Now we consider the electromagnetic field coupled to the time-independent external current $j=(j_0,0)$, where $j_0$ is smooth and
compactly supported. 
The Dirac procedure of quantization with constraints applied to this theory in the
Coulomb gauge gives the familiar formulas for the electromagnetic fields (see e.g. \cite{Wi95})
\begin{align}
\vE_{\mrm{C}}(\vx)&:=  \vE_{\free}(\vx){ +} \fr{1}{\De} \nabla j_0(\vx), \label{C-electric}\\
\vB_{\mrm{C}}(\vx)&:= \vB_{\free}(\vx).  \label{first-magnetic-equality}
\end{align}
 The same quantization
procedure applied in the axial gauge for the axis direction $e=(0,\ve)$ gives instead   (see e.g. \cite{HL94}) 
\begin{align}
\vE_{\ve}(\vx)&:= \vE_{\free}(\vx){ +}\fr{\ve}{\ve\cdot \nabla+0} j_0(\vx), \label{Ax-electric}\\
\vB_{\ve}(\vx)&:= \vB_{\free}(\vx). \label{magnetic-equality}
\end{align}
The Dirac quantization procedure is ambiguous here, as the constraint matrix has many inverses corresponding to 
various regularizations of the singularity in (\ref{Ax-electric}). The choice of $+0$ is natural as it  corresponds to a string-like 
localized electromagnetic potential in the axial gauge \cite{MSY06}.   (A different derivation of (\ref{Ax-electric}),(\ref{magnetic-equality}) will be given in 
Section~\ref{large-gauge-transformations}). The electromagnetic fields are operator-valued distributions
and we denote by 
\beqa
E_{\mrm{C}}(\vf):=\int d^3\vx\,  \vE_{\mrm{C}}(\vx)\cdot \vf(\vx)
\eeqa
 the smearing with a {\bc $\real^3$-valued}, smooth, compactly supported function $\vf$   (and analogously for the remaining quantities). The smeared
fields are self-adjoint, unbounded operators and to avoid the discussion of domain questions we proceed to their bounded functions
$\exp{i \vE_{\mrm{C}}(\vf)}$.  Now we are ready to state and prove our main result:
%%%%%%%%%%%%%%%%%%%%%%%%%%%%
\bet\label{inequivalence-theorem} Suppose that $q:=\int d^3\vx\, j_0(\vx)\neq 0$. Then there is no unitary $U$ on the Fock space  $\mcF$  such that 
\beqa
U\exp{i (\vE_{\mrm{C}}(\vf_{\mrm{el}})+ \vB_{\mrm{C}}(\vf_{\mrm{m}} )  ) }U^*=\exp{i (\vE_{\ve }(\vf_{\mrm{el}})+ \vB_{\ve}(\vf_{\mrm{m}} )  ) } \label{first-inequivalence}
\eeqa
for all smearing functions $\vf_{\mrm{el}}, \vf_{\mrm{m}}$ and some fixed unit vector $\ve\in \real^3$. (The statement remains valid
if we restrict attention to smearing functions  $\vf_{\mrm{m}}\equiv 0$ and $\vf_{\mrm{el}}$ supported in any fixed string 
$\{\, \vx=|\vx| \hat\vx\,:\, |\vx|\in \real_+, \hat\vx \in {\bc \om}\,\}$,
where ${\bc \om}$ is a small neighbourhood of a point on the unit sphere).

Given two unit vectors $\ve\neq \ve'$ there is no unitary $V$  on $\mcF$ s.t.
\beqa \label{axial-shift}
V\exp{i (\vE_{\ve}(\vf_{\mrm{el}})+ \vB_{\ve}(\vf_{\mrm{m}} )  ) }V^*=\exp{i (\vE_{\ve' }(\vf_{\mrm{el}})+ \vB_{\ve'}(\vf_{\mrm{m}} )  ) }
\eeqa
for all smearing functions $\vf_{\mrm{el}}, \vf_{\mrm{m}}$.
\eet
%%%%%%%%%%%%%%%%%%%%%%%%%%%%%%%%%%%%
{\bcol We remark that the assumption $q\neq 0$ is essential. Currents with $q=0$, for which different gauges are unitarily equivalent, are in abundance. 
This can easily be seen from  the  discussion in the next section.}
%%%%%%%%%%%%%%%%%%%%%%%%%%%%%%%%%%%%%

We will prove the theorem by adapting the method of central sequence  to the situation at hand (cf. \cite{Ku98, CD18}).
Let us first justify (\ref{first-inequivalence}). Since it is clear from (\ref{magnetic-equality}) that the problem of inequivalence is related to
the electric field, we can set $\vf_{\mrm{m}}=0$. As  $\vf_{\mrm{el}}$ we choose
\beqa
\vf_{\mrm{el}, \vn, r}(\vx)=   \vn \fr{1}{r} f\left(\fr{\vx-\vn r}{r}\right),
\eeqa
where $f$ is supported in a ball around the origin of radius much smaller than one   and $\vn$ is a unit vector in $\real^3$. With this choice we have
\beqa
\vE_{\mrm{C}}(\vf_{\mrm{el},  \vn, r  })=    \int d^3\vx\, f(\vx)\, r^2\vn\cdot \vE_{\mrm{C} }((\vx+\vn)r   ), 
\eeqa
which is a smeared version of the flux (\ref{flux}). It will be important in the later part of the proof that the following expression 
is independent of $r$
\beqa
 \lan 0| e^{i \vE_{\free }(\vf_{\mrm{el},\vn,r  }   ) } |0\ran= e^{- \fr{1}{4}\int d^3\vk\,  |\vk| |P_{\mrm{tr}}\vn \ti{f}(\vk)|^2},
\eeqa
where  $P_{\mrm{tr}}$ is the transverse projection {\bcol and tilde denotes the Fourier transform}. 
%We proceed by the method of central sequences \cite{}: 

Now we assume by contradiction that  there exists a unitary $U$ as in the theorem  and we compare
the vacuum expectation values of the resulting equality
\begin{align}
\lan 0|U e^{i \vE_{\free}(\vf_{\mrm{el}, \vn,r })  }U^*|0\ran e^{ i \fr{1}{\De} \nabla j_0( \vf_{\mrm{el}, \vn,r } )} = 
\lan 0| e^{i \vE_{\free}(\vf_{\mrm{el},\vn,r  }   ) } |0\ran e^{i\fr{\ve}{\ve\cdot \nabla+0} j_0( \vf_{\mrm{el}, \vn,r }  ) }.
\end{align}
Since the $C^*$-algebra generated by the {\bc free} electromagnetic fields acts irreducibly on $\mcF$,
we can find, by the Kadison  transitivity theorem   \cite[Theorem 10.2.1]{KR}, a unitary $\ti{U}$ in this algebra s.t. $U^*|0\ran=\ti{U}^*|0\ran$.
Then, by locality, $\lim_{r\to\infty} [e^{i \vE_{\free }(\vf_{\mrm{el}, \vn,r })},\ti{U}^*]=0$ in norm and we can write
\begin{align}\label{contradiction}
e^{i \fr{1}{\De} \nabla j_0( \vf_{\mrm{el}, \vn,r } )} + o(r^{-1})= e^{i\fr{\ve}{\ve\cdot \nabla+0} j_0( \vf_{\mrm{el}, \vn,r }  ) },
\end{align}
where $o(r^{-1})$ denotes a term which tends to zero as $r\to\infty$.

Thus to  conclude the proof of {\bcol the first part of the theorem,} %(\ref{first-inequivalence})
 we have to show that the contributions to the flux coming from
the $c$-number parts in (\ref{C-electric}) and (\ref{Ax-electric}) are different. Concerning the Coulomb part, we have
\begin{align}
\fr{1}{\De} \nabla j_0( \vf_{\mrm{el}, \vn,r } ) &= \int d^3\vx f(\vx)\int  d^3\vy \fr{r^2 \vn\cdot ( \vn r +\vx r-\vy)}{4\pi | \vn r+\vx r-\vy|^3}  j_0(\vy)\non\\
&\underset{r\to\infty}{\to} \int d^3\vx f(\vx)\fr{q \vn\cdot (\vn+\vx) }{4\pi |\vn+\vx|^{\bcol 3} }. \label{Coulomb-flux}
\end{align}
We note in passing that in the limit of no smearing ($f(\vx)\to \de(\vx)$) we obtain a spherically symmetric distribution $\fr{q}{4\pi}$ which manifestly respects  the Gauss Law.  
As for the axial part, we set $\vn_{\vx}:=\vn+\vx$  
and compute
 \begin{align}
  \fr{\ve}{\ve\cdot \nabla+0} j_0( \vf_{\mrm{el}, \vn,r }  )   &=\int d^3\vx f(\vx) \lim_{\veps\to 0}\fr{r^2\vn\cdot\ve}{\ve\cdot \nabla_{\vy}+\veps} j_0(\vy)|_{\vy=r\vn_{\vx}} \non\\
&=\int d^3\vx f(\vx)  
(r^2\vn\cdot\ve)  \lim_{\veps\to 0}\int_0^{\infty} ds \,e^{ -(\ve\cdot \nabla_{\vy}+\veps)s} j_0(\vy)|_{ \vy=r\vn_{\vx} } \non\\
&= \int d^3\vx f(\vx)  
(r^2\vn\cdot\ve) \int_0^{\infty} ds\,  j_0(r\vn_{\vx}-s\ve ) \label{intermediate-expression}\\
&= \int d^3\vx f(\vx)  
(  r^2 \vn\cdot\ve) \int_0^{\infty} ds\,  j_0((r (\vn_{\vx}\cdot \ve) -s)\ve+ r P_{\ve}^{\bot}  \vn_{\vx}   )\non\\
&= \int d^3\vx f(\vx)  
(  r^2\vn\cdot\ve) \int_{-r(\vn_{\vx}\cdot \ve) }^{\infty} ds\,  j_0(  -s\ve+ r P_{\ve}^{\bot}\vn_{\vx}   ),
\end{align}
where $P_{\ve}^{\bot}=1-|\ve\ran \lan \ve|$. 
Suppose first that $\vn$ is not parallel to $\ve$. Then, if the support of $f$ is in a sufficiently small neighbourhood of zero, we have
 $ P_{\ve}^{\bot} \vn_{\vx}\neq 0$ and $j_0(  -s\ve+ r P_{\ve}^{\bot}  \vn_{\vx}   )=0$ for $ r$ sufficiently large and all $s$.
 Next, suppose that $\vn=-\ve$. Then, again for a sufficiently small support of $f$, we have that $(\vn_{\vx}\cdot \ve)<0$ and the
 expression  vanishes for $r$ sufficiently large due to the shrinking of the region of $s$-integration. Finally we consider the case
 $\vn=\ve$. Assuming that $\ve$ is in the direction of the third axis of the coordinate frame, we can rewrite expression~(\ref{intermediate-expression})  
as follows
\begin{align}
(\ref{intermediate-expression})&=\int d^3\vx\, f(\vx) r^2\int_0^{\infty}ds\, j_0(rx_1, rx_2, (r-s)+rx_3) \non\\
&= \int d^3\vx\, f(x_1,x_2,x_3) r^2\int_{-\infty}^{ r(1+x_3) }ds'\, j_0(rx_1, rx_2, s')\non\\
&= \int dy_1 dy_2 dx_3\, f(y_1/r,y_2/r,x_3) \int_{-\infty}^{ r(1+x_3) }ds'\, j_0(y_1, y_2, s')\non\\
&\underset{r\to\infty}{\to} q \int  dx_3\, f(0,0,x_3). 
\end{align}
Thus, summing up, we have
\beqa\label{axial-flux}
 \fr{\ve}{\ve\cdot \nabla+0} j_0( \vf_{\mrm{el}, \vn,r }) = \left\{ \begin{array}{ll}
 q \int  ds\, f(s\ve)  & \textrm{for $\vn=\ve$,}\\
0 & \textrm{for $ \vn\neq \ve$. }
\end{array} \right.
\eeqa
We note as an aside that the first line in (\ref{axial-flux}) is singular in the limit $f(\vx)\to \de(\vx)$, thus there is no conflict with the Gauss Law here.
By comparing the second line of (\ref{axial-flux}) with (\ref{Coulomb-flux}) we easily obtain a contradiction in (\ref{contradiction}) in the limit $r\to\infty$.
The second part of the theorem is proven analogously, making use of the first line in (\ref{axial-flux}).
%%%%%%%%%%%%%%%%%%%%%%%%%%%%%%%
\section{Large gauge transformations and Wilson loops} \label{large-gauge-transformations}
%%%%%%%%%%%%%%%%%%%%%%%%%%%%%%%
\setcounter{equation}{0}

In this section we establish relations between different gauges which are less restrictive than
unitary equivalence and thus not in conflict with  Theorem~\ref{inequivalence-theorem}.
We first introduce a general class of  gauge transformations of the  external current QED, {\bcol which is initially} in the Coulomb gauge.
{\bcol Then we} identify the transformation mapping the Coulomb gauge into the axial gauge theory  for a given direction of the axis.

We recall that the Coulomb gauge electromagnetic potential and the Hamiltonian have the form
\newcommand{\vz}{\pmb{z}}
\begin{align}
&A_{0, \mrm{C}}(t,\vx)
:=-\fr{1}{\De} j_0(\vx), \quad  
\vA_{\mrm{C}}(t,\vx):=\vA_{\bot}(t,\vx),\label{em-potential-one}\\
&H_{\mrm{C}}:= H_{\mrm{fr}}+\h\int d^3\vx \,A_{0, \mrm{C}}(\vx) j_0(\vx), \textrm{ where } H_{\mrm{fr}}:= \sum_{\la=\pm} \int d^3\vk\, |\vk|a^*_{\la}(\vk)a_{\la}(\vk).
\end{align}
The resulting electromagnetic fields are given by  (\ref{C-electric}), (\ref{first-magnetic-equality}). While the minimal coupling is not manifest from the
above formulas (note that $H_{\mrm{C}}$ depends quadratically on $j_0$), {\bcol its remnant} is the following relation
\beqa
\fr{\de H_{\mrm{C}}}{\de j_0(\vx)} =A_{0, \mrm{C}}(\vx). \label{minimal-coupling}
\eeqa

To change the gauge we introduce a family of   operator-valued distributions  $\vx\mapsto \chi_{\veps}(\vx)$ s.t. $[ \chi_{\veps}(\vx),   \chi_{\veps}(\vx')]=0$
{\bcol for $\vx, \vx'\in \real^3$}. They
 depend on a regularization parameter $\veps>0$,  whose role will become clear in the example below, and which eventually will  tend to zero.  For non-zero 
 $\veps$, and  after smearing with real-valued  test functions, these distributions  are assumed to
 yield self-adjoint operators.
 Thus we can define the unitaries
\beqa
W_{\veps}:=e^{ -i\chi_{\veps}(j_0)}.  \label{unitaries}
\eeqa   
Setting $\chi_{\veps}(t,\vx):=e^{iH_{\mrm{C}}t } \chi_{\veps}(\vx)     e^{-iH_{\mrm{C}}t } $, we define a new potential as follows
\begin{align}
A_{\veps,  0}(t,\vx)&:=  W_{\veps}( A_{0,\mrm{C}}(t,\vx)   +\pa_t \chi_{\veps}(t,\vx) )  W_{\veps}^*, \label{transformation-law-one}\\
\vA_{\veps}(t,\vx)&:= W_{\veps}( \vA_{\mrm{C}}(t,\vx)   - \nabla \chi_{\veps}(t,\vx) )  W_{\veps}^*.  \label{transformation-law} 
\end{align}
Clearly, the resulting electromagnetic fields
\begin{align}
\vE_{\veps}(t,\vx)&=-\pa_t\vA_{\veps}(t,\vx)-\nabla A_{0,\veps}(t,\vx)=    W_{\veps} \vE_{\mrm{C}}(t,\vx)  W_{\veps}^*,  \label{primed-E}\\
 \vB_{\veps}(t,\vx)&=\nabla \times  \vA_{\veps}(t,\vx)=  W_{\veps} \vB_{\mrm{C}}(t,\vx)  W_{\veps}^*, \label{primed-B}
\end{align}
satisfy the Maxwell equations with the same current $j$.  Their time-evolution is governed by the Hamiltonian
\beqa
H_{\veps}:= W_{\veps} H_{\mrm{C}} W_{\veps}^*. \label{Hamiltonian}
\eeqa 

  The presence of the transformation $W_{\veps}(\ldots)W_{\veps}^*$ in (\ref{transformation-law-one}), (\ref{transformation-law}) calls for a
justification. We remark that the Maxwell equations are insensitive to this transformation in our external current situation\footnote{For Pauli-Fierz 
type models with a dynamical electron the quantum Maxwell equations depend on the electromagnetic potential via the electron's velocity (see e.g.  
\cite[formula (13.51)]{Sp}). Then the form-invariance of the Maxwell equations necessitates  a transformation analogous to $W_{\veps}(\ldots)W_{\veps}^*$
in the definition of gauge transformations.}.
The same is true for the  Dirac brackets, as they only fix the algebraic relations between the potentials and the fields, and not the representation of the resulting algebra. 
However, global quantities, like the Hamiltonian, may be sensitive to the choice of the representation. It is therefore not a surprise that the role
of the transformation   $W_{\veps}(\ldots)W_{\veps}^*$ is to preserve the   minimal coupling property (\ref{minimal-coupling}), as can be seen
from the following computation:
\beqa
\fr{\de H_{\veps}}{\de j_0(\vx)} = e^{-i\chi_{\veps}(j_0)}   \fr{\de H_{\mrm{C}}}{\de j_0(\vx)}   e^{i\chi_{\veps}(j_0)}+ e^{-i\chi_{\veps}(j_0)} [-i\chi_{\veps}(\vx),   H_{\mrm{C}}]
e^{i\chi_{\veps}(j_0)}
=A_{\veps,0}(\vx).
\eeqa

Let us now focus on the transformation from the Coulomb to axial gauge. By imposing the axial gauge condition $\ve\cdot \vA_{\veps}(\vx)\to 0$ as $\veps\to 0 $ we read off from~(\ref{transformation-law})
\beqa
\chi_{\ve,\veps}(\vx)=\fr{1 }{\ve\cdot \nabla  -\veps}\ve \cdot \vA_{\mrm{C}}(\vx). \label{axial-chi}
\eeqa
 We remark that the choice of the $-\veps$ prescription will prove consistent with the choice made in (\ref{Ax-electric}).
 The corresponding family of  unitaries (\ref{unitaries}), similar to transformations considered in \cite{HL94},  has the form 
\beqa
W_{\ve, \veps}:=\exp(i  \int_0^{\infty} ds\, e^{-\veps s}  (\ve\cdot \vA_{\mrm{C}})(j_0)(s\ve) ), \label{transformation-between-gauges}
\eeqa
{\bcol where we used  $(\ve\cdot \nabla  -\veps)^{-1}=\int_0^{\infty} ds\, e^{  (\ve\cdot \nabla  -\veps)s }$}. 
As we show in Lemma~\ref{LGT-proposition} below, the resulting electromagnetic fields (\ref{primed-E}), (\ref{primed-B})
coincide with the axial gauge electromagnetic fields (\ref{Ax-electric}), (\ref{magnetic-equality}) from the previous section. This lemma
should be compared with  the first part of Theorem~\ref{inequivalence-theorem}.  
The key point here is
that $W_{\ve, \veps}$ does not converge to a well-defined unitary in the limit $\veps\to 0$ if $q\neq 0$.  (For $q=0$ such a
limiting unitary {\bcol may} exist, and the statement of Theorem~\ref{inequivalence-theorem} {\bcol may not be}  valid). Nevertheless,  $\lim_{\veps\to 0} W_{\ve, \veps}( \,\cdot \,   )W_{\ve, \veps}^*$
does exist and defines a Bogolubov transformation or, in other words, an automorphism of the $C^*$-algebra of {\bc the free} electromagnetic fields.
%%%%%%%%%%%%%%%%%%%%%%%%%%%%%%%%%%%%%%%%%%%%
\bel\label{LGT-proposition} For a fixed unit vector $\ve\in \real^3$ and all  smearing functions $\vf_{\mrm{el}}, \vf_{\mrm{m}}$ we have
\beqa
  \lim_{\veps\to 0}W_{\ve, \veps} \exp{i (\vE_{\mrm{C}}(\vf_{\mrm{el}})+ \vB_{\mrm{C}}(\vf_{\mrm{m}} )  ) }W_{\ve, \veps}^* =\exp{i (\vE_{\ve }(\vf_{\mrm{el}})+ \vB_{\ve}(\vf_{\mrm{m}} )  ) }. \label{LGT}
\eeqa
\eel
%%%%%%%%%%%%%%%%%%%%%%%%%%%%%%%%%%%%%
\nin This  lemma is a consequence of the canonical commutation relations
\beqa
\,[A_{\bot,i}(\vx),-E_{\free,j}(\vx')]= i\de^{\bot}_{i,j}(\vx-\vx'):=i(2\pi)^{-3} \int d^3\vk\, e^{-i\vk \cdot(\vx-\vx')} (\de_{i,j}-\hat{k}_i \hat{k}_j) \label{CCR-transverse}
\eeqa
and the following computation
\beqa
& &W_{\ve, \veps} E_{\free,j}(\vx) W_{\ve, \veps}^*- E_{\free,j}(\vx)= [i   \int_0^{\infty} ds\, e^{-\veps s} (\ve\cdot \vA_{\bot})(j_0)(s\ve), E_{\free,j}(\vx)]\non\\
& &\ph{444444444}= -i   \int_0^{\infty} ds\, \int d^3\vx'\, e^{-\veps s}  ie_i \de_{i,j}^{\bot}(   s\ve+\vx'-\vx  )j_0(\vx')\non\\ 
%& &\ph{444444444}={\bc +}\fr{1}{(2\pi)^{3/2}} \int_0^{\infty} ds\, \int d^3\vk\, e^{-\veps s} e^{-i\vk s \ve+i \vk \cdot \vx}e_i\cdot (\de_{i,j}-\hat{k}_i\hat{k}_j )  \ti{j_0}(\vk) \non\\
& &\ph{444444444}= \fr{1}{(2\pi)^{3/2}}   \int_0^{\infty} ds\, \int d^3\vk\, e^{-i(\vk\cdot \ve-i\veps)s+i\vk\cdot \vx}e_i (\de_{i,j}-\hat{k}_i\hat{k}_j )  \ti{j_0}(\vk)\non\\
%& &\ph{444444444}= {\bc +}\fr{1}{(2\pi)^{3/2}}   \int d^3\vk\, e^{i\vk \vx}\fr{1}{i(\vk \cdot \ve-i\veps)} e_i\cdot (\de_{i,j}-\hat{k}_i\hat{k}_j )  \ti{j_0}(\vk) \non\\
& &\ph{444444444}= \fr{1}{(2\pi)^{3/2}}   \int d^3\vk\, e^{i\vk\cdot \vx}\fr{1}{i(\vk \cdot \ve-i\veps)} e_j   \ti{j_0}(\vk) \label{axial-term-x}\\
& &\ph{44444444444} -\fr{1}{(2\pi)^{3/2}}   \int d^3\vk\, e^{i\vk \cdot \vx}\fr{1}{i(\vk \cdot \ve-i\veps)} \fr{(\ve\cdot \vk)}{|\vk|^2} k_j   \ti{j_0}(\vk).  \label{Coulomb-term-x}
\eeqa
 We recall formulas  (\ref{C-electric}) and (\ref{Ax-electric}),  and  note that 
\beqa
& &(\ref{axial-term-x})=\fr{{\bcol e_j}}{ \ve  \cdot\nabla +\veps} j_0(\vx), \\
& &(\ref{Coulomb-term-x})=   \fr{1}{(2\pi)^{3/2}}   \int d^3\vk\, e^{i\vk \cdot  \vx}  \fr{1}{|\vk|^2} ik_j   \ti{j_0}(\vk)+ o(\veps)\underset{\veps \to 0}{ \to } -\fr{\nabla_{j}  j_0(\vx) }{\Delta}, 
\eeqa
where $o(\veps)$ denotes a term which tends to zero as $\veps\to 0$.
Thus we  conclude the proof of  Lemma~\ref{LGT-proposition},  since  $W_{\ve, \veps}$ commutes with the magnetic fields. \vspace{0.3cm}
%%%%%%%%%%%%%%%%%%%%%%%%%%%%%%%%%%%%%%%%%%
%%%%%%%%%%%%%%%%%%%%%%%%%%%%%%%%%%%%%%%%%%%%%
\begin{figure}
\centering
\begin{subfigure}{.5\textwidth}
  \centering
  \includegraphics[width=7cm]{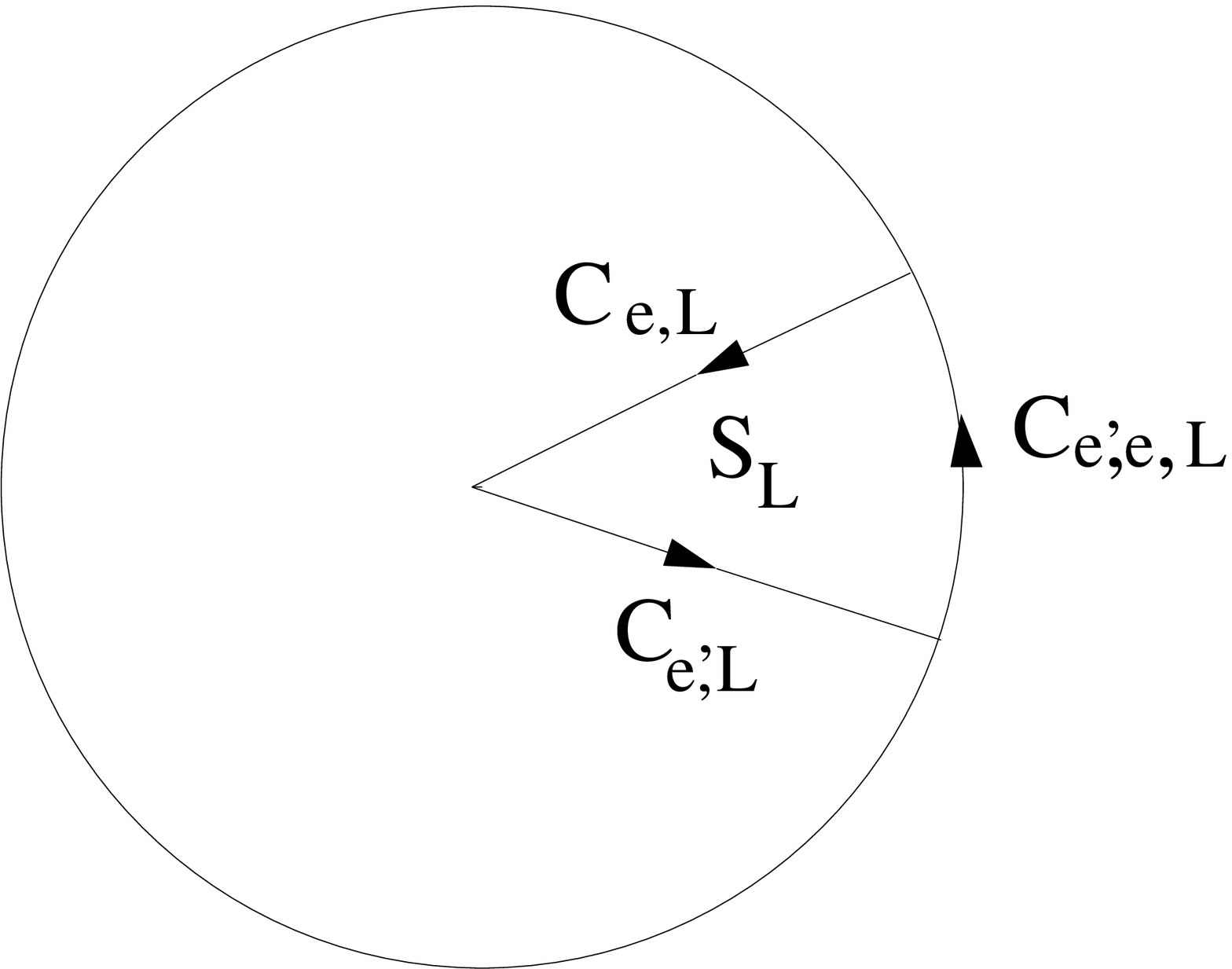}
  \caption*{Fig.~1. Contour of integration in (\ref{contour-integral}).}
  \label{fig:sub1}
\end{subfigure}%
\begin{subfigure}{.5\textwidth}
  \centering
  \includegraphics[width=7.2cm]{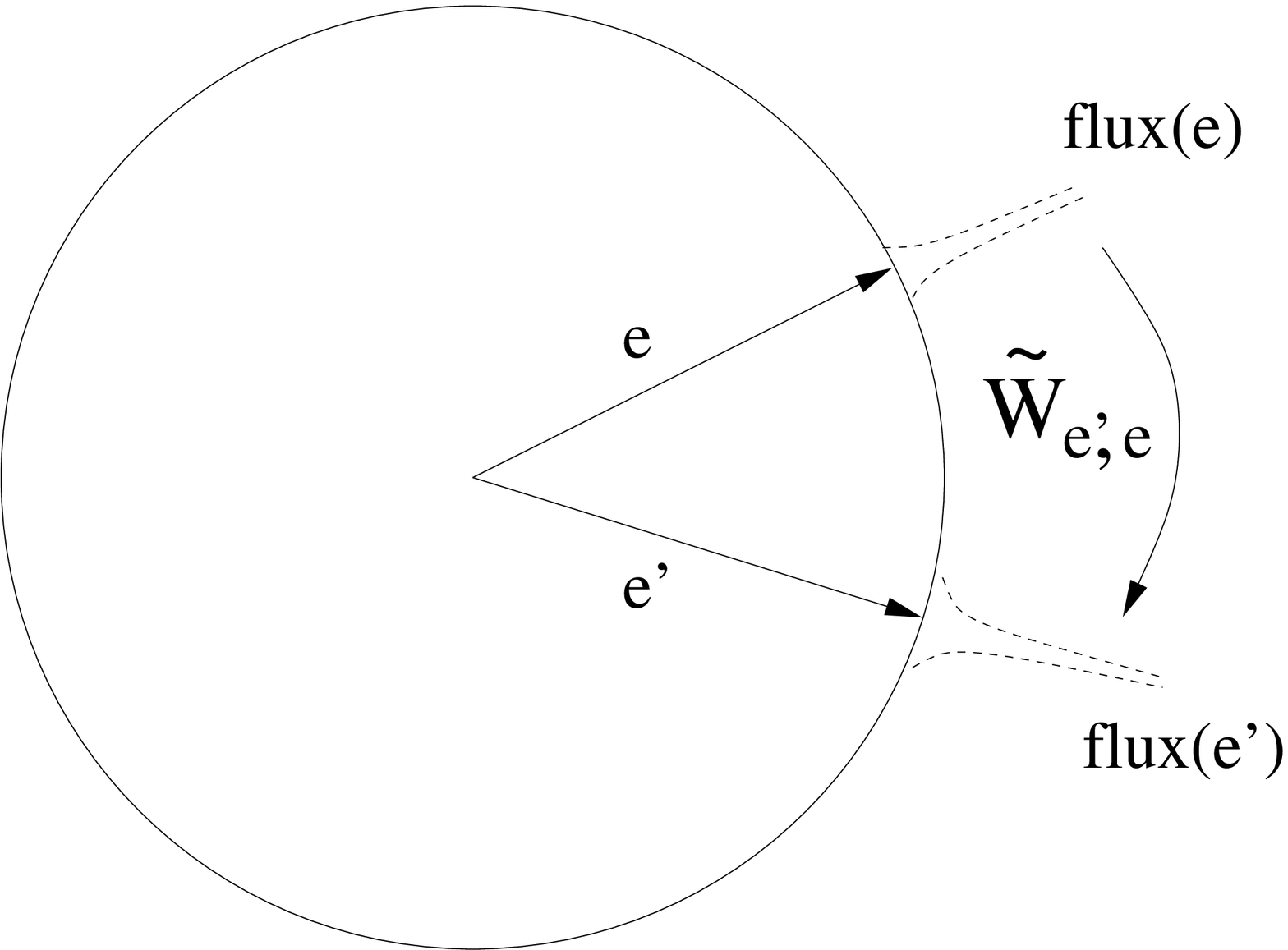}
  \vspace{0.1cm}
    \caption*{Fig.~2. $\wt{W}_{\ve',\ve}$ as a string-local flux-carrying field.}
  \label{fig:sub2}
\end{subfigure}
%\caption{Two pictures: (\ref{fig:sub1})  xxx.  (\ref{fig:sub2}) xxx. }
\label{mass-shell-fig}
\end{figure}
%%%%%%%%%%%%%%%%%%%%%%%%%%%%%%%%%%%%%%

From (\ref{LGT}) we can easily read off the large gauge transformation linking  two axial gauges with axes $\ve\neq \ve'$.  The relation
\beqa
\lim_{\veps\to 0} W_{\ve', \veps} W_{\ve, \veps}^* \exp{i (\vE_{\ve }(\vf_{\mrm{el}})+ \vB_{\ve}(\vf_{\mrm{m}} )  ) } W_{\ve, \veps} W_{\ve', \veps}^* 
=   \exp{i (\vE_{\ve' }(\vf_{\mrm{el}})+ \vB_{\ve'}(\vf_{\mrm{m}} )  ) }  \label{change-of-axis}
\eeqa
should be compared with the second part of Theorem~\ref{inequivalence-theorem}. It turns out that this transformation has interesting geometric
properties. To bring them to light, we change the regularization method. That  is, we define an auxiliary
family of transformations 
\beqa
W_{\ve,L}:= \exp( i  \int_0^{L} ds\,  (\ve\cdot \vA_{\mrm{C}})(j_0)(s\ve))
\eeqa
and  check by a straightforward computation that (\ref{change-of-axis}) remains true if the operators $W_{\ve, \veps},  W_{\ve', \veps}$ are replaced with $W_{\ve,L}, W_{\ve',L}$
and the limit $L\to \infty $ is taken. Clearly, we can write
 \begin{align}
W_{\ve',\ve,L}&:= W_{\ve',  L} W_{\ve,  L}^*
= \exp(i\int_0^{L} ds'\,  (\ve'\cdot \vA_{\mrm{C}})(j_0)(s'\ve')+i\int_{-L}^{0} ds\, ((-\ve)\cdot \vA_{\mrm{C}})(j_0)(s(-\ve)) ).
\end{align}
Let us denote the two  regions of integration above by $C_{\ve, L}, C_{\ve',L}$. Aiming at a Wilson loop,
 we  close  $C_{\ve, L}\cup C_{\ve',L}$   with a contour  $C_{\ve',\ve,L}$ depicted in Fig. 1. We set $\pa S_L:= C_{\ve, L}\cup C_{\ve',L}\cup  C_{\ve',\ve,L}$
and define  a new transformation
\beqa
\wt{W}_{\ve',\ve,L}:= \exp(i\int_{\pa S_L} \vA_{\mrm{C}}(j_0)(\vr)\cdot d\vr)=\exp(i\int_{S_L} \vB_{\mrm{C}}(j_0)(\vr)\cdot d\boldsymbol{S} ),  \label{contour-integral}
\eeqa 
where in the second step we used the Stokes law and $S_L$ is the surface enclosed by $\pa S_L$.  Since  $\vB_{ \mrm{C}  }$ is a local field,  we see that the resulting Bogolubov transformation acts trivially on all the observables  spacelike separated w.r.t. the region $\bigcup_{L\geq 0}  S_L+\supp j_0$. Due to the following
theorem, closing the contour of integration has no effect in the limit $L\to \infty$. Thus the operation of  changing the  axial gauge  from $\ve$ to $\ve'$ is
localised in the string defined by the two axes (see Fig.~2).  In the light of formula~(\ref{axial-flux}), this operation has the physical meaning of carrying the  flux of the electric field (\ref{flux}) from one axis direction to another.  This observation may be relevant for a development of  superselection theory for local gauge invariance
in the spirit of DHR \cite{Ha}.
%%%%%%%%%%%%%%%%%%%%%%%%%%%%%%%%%%
\bet For any fixed unit vectors $\ve\neq \ve'$ and all  smearing functions $\vf_{\mrm{el}}, \vf_{\mrm{m}}$ we have
\beqa
\lim_{L \to \infty}\wt{W}_{\ve',\ve,L} \exp{i (\vE_{\ve }(\vf_{\mrm{el}})+ \vB_{\ve}(\vf_{\mrm{m}} )  ) }  \wt{W}_{\ve',\ve,L}^*=   \exp{i (\vE_{\ve' }(\vf_{\mrm{el}})+ \vB_{\ve'}(\vf_{\mrm{m}} )  ) }. 
\eeqa
\eet
%%%%%%%%%%%%%%%%%%%%%%%%%%%%%%%%%%
\nin Clearly, it suffices to show that the contribution to $\lim_{L\to\infty}\wt{W}_{\ve',\ve,L}(\,\cdot\,)\wt{W}_{\ve',\ve,L}^*$  coming from the  contour  
$C_{\ve',\ve,L}$ acts as the identity on the electromagnetic fields. As it is trivially true for the magnetic fields, it suffices to show that
\beqa
\lim_{L\to\infty}\exp(i\int_{C_{\ve',\ve,L}} \vA_{\mrm{C}}(j_0)(\vr)\cdot d\vr) e^{i\vE_{\free}(\vfe)} \exp(  - i\int_{C_{\ve',\ve,L}} \vA_{\mrm{C}}(j_0)(\vr)\cdot d\vr)=  e^{i\vE_{\free}(\vfe)}.\eeqa
For this purpose we compute, using the canonical commutation relations (\ref{CCR-transverse}),
\beqa
& &\,[\int_{C_{\ve,\ve',L}} \vA_{\bot}(j_0)(\vr)\cdot d\vr,  -E_{\free,j}(\vfe)]\non\\
& &\ph{444}=\int d^3\vx \int d^3\vy \, j_0(\vx) (\vfe)_j(\vy) \int_{  C_{\ve',\ve,L}   } (d\vr)_i  [A_{\bot,i}(\vx+\vr), -\E_{\free,j}(\vy)] \non\\
& &\ph{444}=i\int d^3\vx \int d^3\vy \, j_0(\vx) (\vfe)_j(\vy) \int_{   C_{\ve',\ve,L}  } (d\vr)_i (2\pi)^{-3} \int d^3\vk\, e^{-i\vk \cdot(\vx+\vr-\vy)} (\de_{i,j}-\hat{k}_i \hat{k}_j)  \non\\
& &\ph{444}=i\int_{C_{\ve',\ve,L}} (d\vr)_i  \int d^3\vk\, e^{-i\vk\cdot  \vr} (\de_{i,j}-\hat{k}_i \hat{k}_j)\ti{j}_0(\vk) (\ti{\vf}_{\mrm{el}})_j(-\vk)  \non\\
 & &\ph{444}= i\int_{C_{\ve',\ve,L}} (d\vr)_i \de_{i,j} \int d^3\vx\, j_0(\vx) (\vfe)_j(\vx+\vr)\label{first-AB}\\
& & \ph{44444}+i\int_{  C_{\ve',\ve,L}  } (d\vr)_i  \int d^3\vx\, j_0(\vx)  \int d^3\vy\, \fr{1}{4\pi|\vy+\vr|} \pa_i\pa_j(\vfe)_j(\vx-\vy). \label{second-AB} 
\eeqa
We note that (\ref{first-AB}) vanishes for sufficiently large $L$ since $j_0$ and $\vfe$ are compactly supported. As for (\ref{second-AB}),
we write in polar coordinates $\vr=L\hat{\vr}$  and assume that the contour is in the plane $\{x_3=0\}$. Then $d\vr=L   \pmb{\hat{\varphi}} d\varphi$
and we have
\beqa
(\ref{second-AB})= i\int_{0}^{\varphi_0} d\varphi \,  \hat{\pmb{\varphi}}_i \int d^3\vx\, j_0(\vx)  \int d^3\vy\, \fr{L}{ 4\pi|\vy+L \hat{\vr} |} \pa_i\pa_j(\vfe)_j(\vx-\vy).
\eeqa
We take  the limit $L\to \infty$ and then the integral w.r.t. $\vy$ gives zero.

%%%%%%%%%%%%%%%%%%%%%%%%%%%%%
\section{Conclusion and outlook}\label{conclusions}
%%%%%%%%%%%%%%%%%%%%%%%%%%%%%%%
\setcounter{equation}{0}

In this paper we demonstrated the unitary inequivalence of  different  gauge-fixing conditions  in the presence of a non-zero electric charge.
This was achieved by exhibiting an asymptotic charge which distinguished different
gauges. Furthermore, we presented a general formalism for computing large gauge transformations
linking  different gauges. We showed that the transformation pertaining
to the change of the  axis direction {\bc of } the axial gauge is given by a Wilson loop over the region 
confined by the two axes.  Although our analysis was restricted to the external current situation, 
we believe that the main conclusions remain valid in a larger generality. In particular, it is very plausible that 
the flux of the electric field (\ref{flux}) retains the symmetry of the gauge fixing condition  also in the presence 
of dynamical charged particles.

 {\bc It is well known  that the axial gauge is very singular. In our case}, the Hamiltonian
(\ref{Hamiltonian}) exists in the limit $\veps\to 0$ only as a quadratic form and a presence of a self-adjoint Hamiltonian
can be excluded using criteria from \cite{Ro70}.  Similarly, $A_{0, \veps}$ diverges in the limit $\veps\to 0$.  
These problems can be resolved by considering smeared axial gauges in the spirit of \cite{MSY06}. This  amounts to replacing (\ref{axial-chi}) with
\beqa
\chi_{g,\veps}(\vx)=\int d\Om(\ve) g(\ve)\fr{1 }{\ve\cdot \nabla -\veps}\ve \cdot \vA_{\mrm{C}}(\vx), \label{smeared-gauge-fixing-condition}
\eeqa
 where  $d\Om$ is the spherical measure and $g$ is a smooth, positive function, normalized to one on the sphere $S^2$. 
 For this choice of $\chi$ the Hamiltonian and the electromagnetic potential are well defined. It is not straightforward
 to relate  the smeared axial gauge to the method of Dirac brackets, but it is possible \cite{We18}.  The discussion from the present paper
can be adapted to the gauges of the form (\ref{smeared-gauge-fixing-condition}). In particular, for $g\equiv C_g\neq 0$ on a subset $O_g\subset S^2$ and
$g\equiv 0$ outside of some larger set $\ti O_g$ we have the following counterpart of formula (\ref{axial-flux}) 
\beqa\label{smeared-gauge}
\lim_{r\to\infty} \int d\Om(\ve) g(\ve)  \fr{\ve}{\ve\cdot \nabla+0} j_0( \vf_{\mrm{el}, \vn,r }) = \left\{ \begin{array}{ll}
  C_g \int d^3\vx f(\vx)\fr{q \vn\cdot (\vn+\vx) }{4\pi |\vn+\vx|^{\bcol 3} }  & \textrm{for $\vn\in O_g$,}\\
0 & \textrm{for $ \vn\notin \ti O_g$. }
\end{array} \right.
\eeqa
It is obtained by similar steps as (\ref{axial-flux}) and making use of the simple fact that for $g\equiv 1/(4\pi)$
formula~(\ref{smeared-gauge-fixing-condition}) reproduces the Coulomb gauge. We note as an aside that the
limit $r\to\infty$ in (\ref{smeared-gauge}) cannot be interchanged with the integral over $d\Om$, as then the result would be zero
by (\ref{axial-flux}).  Exploiting (\ref{smeared-gauge}), one easily shows the inequivalence of  different smeared axial gauges,
for suitable choices of smearing functions,  in analogy with Theorem~\ref{inequivalence-theorem}. Furthermore, 
for a generic non-constant $g$  the Hamiltonian  $H_g$   in the smeared axial gauge does not have a ground state (cf. \cite[formula (32)]{BHS63}).
In this case there is clearly no unitary $U$ s.t. $UH_gU^*=H_{\mrm{C}}$, since $H_{\mrm{C}}$ has a ground state. Summing up, the case of smeared axial gauges
demonstrates that  Theorem~\ref{inequivalence-theorem} is not just a manifestation of the known pathologies of the sharp 
axial gauge.

It is an interesting question for future research how to reconcile the  results of the present paper
with the generally expected gauge independence. One approach is to immerse the system in
a highly fluctuating but low-energetic background radiation (`infravacuum') and try to  restore the unitary equivalence of different gauges.  Concrete examples of 
 such infravacua can be found in \cite{KPR77, Ku98, CD18}. Another approach is to restrict attention to observables
localised in a fixed future lightcone, so that the fluxes (\ref{flux}) cannot be measured.  Then one can try to 
prove the unitary equivalence on the resulting subalgebra,  building on ideas  from \cite{BR14, CD19}. We hope to come back to these questions
in future investigations. 
  
  \vspace{0.2cm}

\nin\textbf{Acknowledgements:} W.D. would like to thank Yoh Tanimoto and Kasia Rejzner for numerous interesting discussions in the course of this work.
The hospitality of the Perimeter Institute,  Waterloo, and the Hausdorff  Research Institute for Mathematics, Bonn, is gratefully acknowledged.   
Thanks are also due to the organizers of the conferences  `Infrared Problems in QED and Quantum Gravity', Perimeter Institute, Waterloo, and 
 `Foundational and Structural Aspects of Gauge Theories', MITP Mainz, which stimulated this research. %We would like to thank Henning Bostelmann, Detlev Buchholz, Fabio Ciolli, Maximilian Duell, Simon~Ruijsenaars and Yoh~Tanimoto for discussions


\begin{thebibliography}{99} 

\bibitem[BHS63]{BHS63} H.-J. Borchers, R. Haag and B. Schroer. \emph{The vacuum state in quantum field
theory}. Nuovo Cimento \textbf{29}, (1963) 148--162.


\bibitem[BCRV19]{BCRV19} D. Buchholz, F. Ciolli, G. Ruzzi and E. Vasselli. \emph{On string-localized
potentials and gauge fields.}  \verb+arXiv:1904.10055+.

\bibitem[Bu82]{Bu82} D. Buchholz.
	\emph{The physical state space of quantum electrodynamics}.
	Commun. Math. Phys. \bf 85\rm, (1982) 49--71. 
	
\bibitem[BR14]{BR14} D. Buchholz and J.E. Roberts. \emph{New light on infrared problems: sectors, statistics, symmetries and spectrum.}
	Commun. Math. Phys.  \bf 330\rm, (2014) 935-972.
	




\bibitem[CD18]{CD18} D. Cadamuro and W. Dybalski. \emph{Relative normalizers of automorphism groups, infravacua and the problem of
velocity superselection in QED}. To appear in Commun. Math. Phys. \verb+arXiv:1807.07919+.

\bibitem[CD19]{CD19}  D. Cadamuro and W. Dybalski. \emph{Curing velocity superselection in non-relativistic QED by restriction to a lightcone}.
 \verb+arXiv:1902.09478+.


\bibitem[CL15]{CL15} M. Campigliaa and A. Laddha.  \emph{Asymptotic symmetries of QED and WeinbergÕs soft photon theorem}. JHEP07 (2015) 115.

\bibitem[CE17]{CE17} M. Campiglia and R. Eyheralde. \emph{Asymptotic U(1) charges at spatial infinity}. JHEP11 (2017) 168. 

\bibitem[DDMS10]{DDMS10}  D.-A. Deckert, D. D\"urr, F. Merkl, M. Schottenloher.  \emph{Time evolution of the external field problem in QED.}   
J. Math. Phys. \textbf{51}, (2010) 122301.


\bibitem[FS15]{FS15} F. Finster and  A. Strohmeier. \emph{Gupta-Bleuler quantization of the Maxwell field in globally hyperbolic space-times}. Ann. Henri Poincare \textbf{16} (2015) 1837--1868, erratum Ann. Henri Poincare \textbf{19} (2018) 323--324.


\bibitem[GS16]{GS16} B. Gabai and A. Sever. \emph{Large gauge symmetries and asymptotic states in QED}. JHEP12 (2016) 095

\bibitem[Ha]{Ha} R. Haag. \emph{Local Quantum Physics}. Second edition. Springer-Verlag, Berlin, 1996.


\bibitem[HL94]{HL94}  K. Haller and E. Lim-Lombridas.  \emph{Quantum gauge equivalence in QED}. Found. Phys. \textbf{24},  217--247 (1994).



\bibitem[HMPS14]{HMPS14}   T. He, P. Mitra, A.P. Porfyriadis and A. Strominger.   
\emph{New symmetries of massless QED}. JHEP10 (2014) 112.



\bibitem[He17]{He17} A. Herdegen. \emph{Asymptotic structure of electrodynamics revisited}. Lett. Math. Phys.  \textbf{107}, (2017) 1439Ð1470.  

\bibitem[KR]{KR} R.V. Kadison and J.R. Ringrose. \emph{Fundamentals of the theory of operator algebras: Advanced theory.} Academic Press, 1986.

\bibitem[KPR77]{KPR77} K. Kraus, L. Polley and G. Reents. \emph{Models for infrared dynamics. I. Classical currents.} Ann. Inst. H. Poincar\'e  \textbf{26}, (1977) 109--162.

\bibitem[Ku98]{Ku98}  W. Kunhardt. \emph{On infravacua and the localization of sectors}.    J. Math. Phys. \textbf{39},  (1998) 6353.


\bibitem[MSY06]{MSY06} J. Mund, B. Schroer and J. Yngvason. \emph{String-localized quantum fields and modular localization} 
Commun. Math. Phys. \textbf{268},   621--672 (2006).
%arXiv:math-ph/0511042v2.

\bibitem[MRS19]{RMS19}  J. Mund, K.H. Rehren  and B. Schroer. \emph{GaussÕ law and string-localized quantum field theory}.  
\verb+arXiv:1906.09596+.



\bibitem[NTO94]{Na94} Y. Nakawaki,  A. Tanaka and K. Ozaki.  \emph{Verification of equivalence of the axial gauge to the Coulomb gauge in QED by embedding in the indefinite metric Hilbert space}
Progress of Theoretical Physics \textbf{91},  579--590 (1994).



\bibitem[Ro70]{Ro70} G. Roepstorff. \emph{Coherent photon states and  spectral condition}. Commun. Math. Phys. \textbf{19}, 301-- 314  (1970).

\bibitem[SDH14]{SDH14} K. Sanders, C. Dappiaggi and T.P. Hack. \emph{Electromagnetism, Local Covariance,
the Aharonov--Bohm Effect and GaussÕ Law}.  Commun. Math. Phys. \textbf{328}, 625Ð667 (2014).

\bibitem[Sch19]{Sch19}  {\bcol B. Schroer. \emph{The role of positivity and causality in interactions involving higher spin}. Nucl. Phys. B \textbf{941}, 91--144 (2019)}.

\bibitem[Sp]{Sp} H. Spohn. \emph{Dynamics of charged particles and their radiation fields}.  Cambridge 2004

%\bibitem[St17]{St17} A.~Strominger. \emph{Lectures on the infrared structure of gravity and gauge theory}. Princeton University Press, 2017.

\bibitem[We18]{We18} B. Wegener. \emph{The problem of inequivalence  of different gauges in external current QED}. MSc thesis,
Technische Universit\"at M\"unchen / Ludwig-Maximilians-Universit\"at M\"unchen, 2018. https://www.theorie.physik.uni-muenchen.de/TMP/theses/thesiswegener.pdf 

\bibitem[Wi95]{Wi95} S. Weinberg. \emph{The quantum theory of fields: Volume 1, Foundations}.
Cambridge University Press, 1995.


\end{thebibliography}
\end{document}